\documentclass[pra,amssymb,nobibnotes,showpacs,twocolumn,floatfix]{revtex4}
\usepackage{graphicx,amsfonts,amssymb,amsmath,hyperref}

\begin{document}

\title{Dynamics Reflects Quantum Phase Transition of Rabi Model}

\author{M. Li}
\affiliation{Qingdao University of Technology, 266520, Qingdao, Shandong, China}
\author{Y. N. Wang}
\affiliation{Shandong University, 266237, Qingdao, Shandong, China}
\author{Z. Y. Song}
\affiliation{Qingdao University of Technology, 266520, Qingdao, Shandong, China}
\author{Y. M. Zhao}
\affiliation{Qingdao University of Technology, 266520, Qingdao, Shandong, China}
\author{X. L. $Zhao^*$}
\affiliation{Qingdao University of Technology, 266520, Qingdao, Shandong, China}
\author{H. Y. $Ma^{\dag}$}
\affiliation{Qingdao University of Technology, 266520, Qingdao, Shandong, China}

\date{\today}

\begin{abstract}
As the simplest and most fundamental model describing the interaction between light and matter,
a breakdown in the rotating wave approximation of the Rabi model leads to phase transition versus
coupling strength when the frequency of the qubit greatly surpasses that of the oscillator.
Besides the phase transition revealed in the ground state, we show that the dynamics of physical
quantities can reflect such a phase transition for this model. In addition to the excitation of
the bosonic field in the ground state, we show that the witness of inseparability (entanglement),
mutual information, quantum Fisher information, and the variance of cavity quadrature can be
employed to detect the phase transition in quench. We also reveal the negative impact of
temperature on checking the phase transition by quench. This model can be implemented using
trapped ions, superconducting artificial atoms coupled bosonic modes, and quantum simulations.
By reflecting the phase transition in a fundamental quantum optics model without imposing the
thermodynamic limit, this work offers an idea to explore phase transitions by non-equilibrium
process for open quantum systems.
\end{abstract}
\maketitle
\section{Introduction}\label{Introduction}
The interaction between light and matter, as well as between harmonic oscillators and atoms,
is pervasive in nature and often associated with quantum phase transitions. Quantum phase
transition occurs when a non-thermal parameter scans across a critical point, causing a sudden
and significant change in the ground state properties of the system, often accompanied by
spontaneous symmetry breaking~\cite{Sachdev,RPP662069}. Such phase transitions can be revealed by various
means and have attracted wide attention in quantum information~\cite{RMP80517,PRL90227902}
and quantum metrology~\cite{NatPhoton5203,PRA80012318}.

Quantum phase transitions have emerged as a prominent area of research in the field of condensed
matter physics. Superradiance is one of quantum phase transition occurs when the coupling
strength between the two subsystems exceeds a critical
threshold. The exploration of this phenomenon commenced with the Dicke model, a theoretical
framework that delves into the collective behavior of a multitude of atoms interacting with
a single harmonic-oscillator mode of the electromagnetic field. Within this model, the atoms
demonstrate quantum-coherent collective behavior, leading to a flurry of captivating dynamics
and effects~\cite{PR9399}. Namely, the superradiance phase transition is commonly examined
under the assumption of the thermodynamic limit, and the interaction between natural atoms
and the cavity field is significantly weaker in comparison to the bare atom and cavity
frequencies~\cite{Sachdev}. Recently, significant advancements have been made in
superconducting qubit circuits, leading to the achievement of the highly-anticipated strong
and ultrastrong-coupling regime~\cite{NP6772,PRL105237001,AP16767}. Furthermore, trapped ion
quantum simulation presents an opportunity to replicate a similar model by ascribing the
oscillatory motion as the harmonic degree of freedom~\cite{PRX8021027}. Given that
the artificial atom or ion investigated in these studies play the role as a two-level system,
namely, qubits, the quantum Rabi model (QRM) can showcase behavior that bears resemblance to
a superradiance phase transition~\cite{PRB69113203,PRA81042311,PRA87013826,PRL115180404,PRL118073001,NC121126}.
Then we use quantum phase transition hereafter to represent superradiance phase
transition in this work.

In this work, we first show the ground-state quantum phase transition for the Rabi model by
several physical quantities, such as the energy-level structure, excitations of the qubit
and the cavity field, quantum Fisher information, measures of entanglement (inseparability),
mutual information, and variance of the cavity field quadrature. Notably, we explore these
quantities without the constraint of a thermodynamic limit, but instead by considering a
higher ratio between the eigen-frequency of the qubit and the cavity field. Then going beyond
this equilibrium by quenching the coupling strength across the critical point of the ground
state phase transition, the dynamics of the quantities are examined to indicate the phase.
Temperature should be considered for a cavity field interacting with entangled atom
pairs in the presence of decoherence~\cite{JPC1234035}. We also check the influence of thermal
excitation on the dynamics and propose potential experimental platforms. This comprehensive
investigation propose a method for illuminating the occurrence of phase transition by
non-equilibrium processes.

The organization of this study is as follows: In Sec.~\ref{Model}, we present the Rabi model
with a significant disparity in eigen-frequencies between the qubit and cavity field. In
Sec.~\ref{PhaseDiagram}, we show the phase diagram concerning various quantities versus the
coupling strength and the eigen-frequency with constraint. In Sec.~\ref{Dynamics}, we check
the behavior of these quantities during a quench to observe and characterize the phase
transition. In Sec.~\ref{InfluenceT}, we explore the impact of environmental temperature on
the dynamics. In Sec.~\ref{Experiments}, we propose potential experimental platforms to realize
this work. Finally, we conclude our work in Sec.~\ref{Conclusion}.
\section{The model}
\label{Model}
\begin{figure}[htbp]
\includegraphics[width=0.9\linewidth]{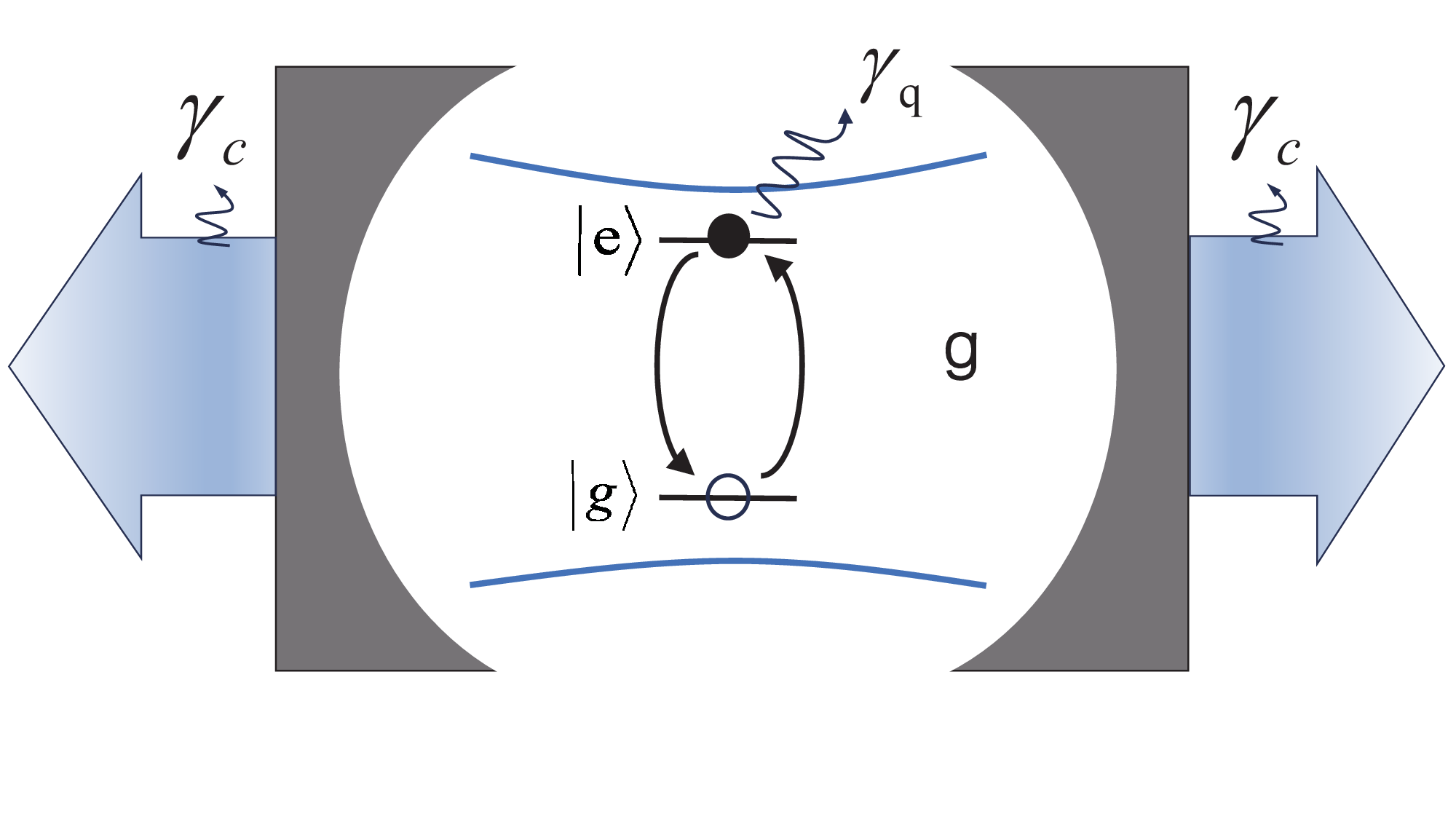}
\caption{Sketch of the quantum Rabi model: a single two-level atom (qubit) coupled to
the cavity field with coupling strength $g$. The ground (excited) state of the atom is
labeled as $|g\rangle$ ($|e\rangle$). Here, $\gamma_c$ and $\gamma_q$ are the cavity
and atomic decay rates, respectively.}\label{fig:fig1}
\end{figure}
The quantum Rabi model~\cite{PR51652,PRA96013849} with a single-mode bosonic field
(such as a cavity mode) coupled to a two-level atom (generic qubit) as depicted in
Fig.~\ref{fig:fig1}, is described by the Hamiltonian ($\hbar=1$ hereafter)
\begin{equation}
\begin{aligned}
\hat{H} =\omega_c \hat{a}^{\dag}\hat{a}+\frac{\omega_q}{2} \hat{\sigma}_z+g(\hat{a}+\hat{a}^{\dag})(\hat{\sigma}_-+\hat{\sigma}_+),
\end{aligned}\label{Eq:Hamiltonian}
\end{equation}
where $\hat{a}^{\dagger}$($\hat{a})$ is the creation (annihilation) operator for the single
mode cavity field with frequency $\omega_c$, and $\hat{\sigma}_{z}$ is the Pauli
z-basis operator with commutation relation $[\hat{\sigma}_+,\hat{\sigma}_-] = \hat{\sigma}_z$.
The parameter $g$ represents the coupling strength between the two subsystems. In the case
of trapped ion systems, the phase-transition-poisonous vector potential item can be neglected
safely~\cite{PRL118073001,PRX8021027,NC121126}.

In the regime of weak coupling ($g\ll\omega_c, \omega_q$), one can simplify the quantum
Rabi model by applying the rotating wave approximation, resulting in Jaynes-Cummings
model~\cite{IEEE5189,JPB38S551,RPP691325} which has been investigated widely in cavity
QED system~\cite{RMP73565}. In scenarios where the coupling strength reaches
or exceeds the magnitude of the frequencies of the cavity mode and qubit, the rotating
wave approximation becomes invalid. This breakdown paves the way for the emergence of
the strong, ultrastrong and even deep-strong coupling regime, facilitating connections
between manifolds characterized by different total excitations~\cite{PRX8021027}.
Many exotic physical properties have been investigated in a plenty of strong coupling
quantum systems such as trapped ions~\cite{PRL118073001,PRX8021027,NC121126}, circuit
QED~\cite{NP6772,NC8779,NP1339,NP1344,NC81715}, and photonic system~\cite{PRL108163601}.

Quantum phase transition is a focus of research usually considered in the thermodynamic
limit since this limit leads to the non-analytic behavior of the free energy or partition
function. However, it was recently realized that a quantum phase transition can also occur
in a small system with only a two-level atom coupled to a bosonic mode, described by the
quantum Rabi model~\cite{PRA87013826}. Since the smaller ratio of the frequencies
$\frac{\omega_c}{\omega_q}$ in the model, the more obviously of the superradiance phase
transition versus the coupling strength, we focus on the situation with $\omega_c \omega_q=1$
in the Hamiltonian~(\ref{Eq:Hamiltonian}).

\begin{figure}[htbp]
\includegraphics[width=1\linewidth]{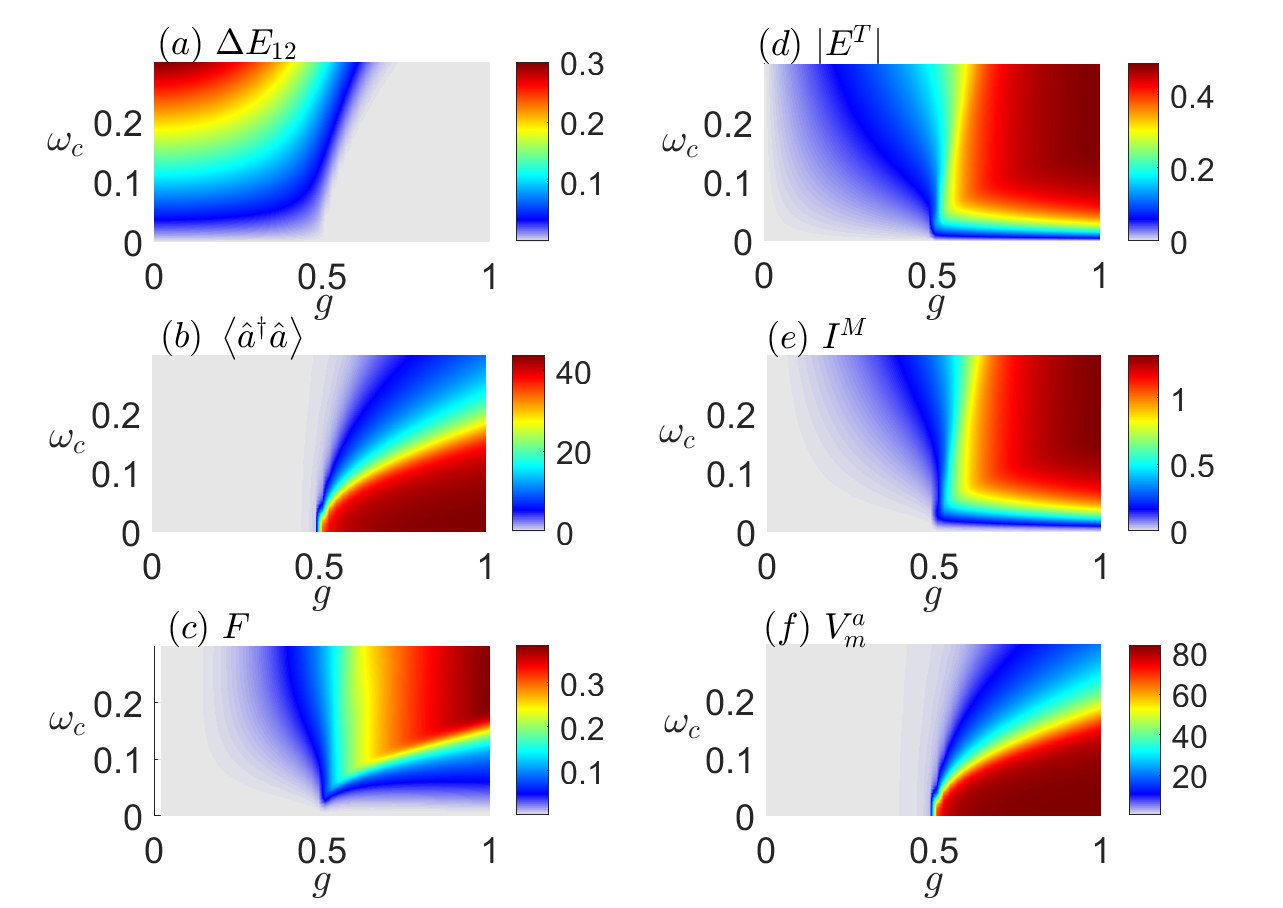}%
\caption{Quantities used to reveal the phase transition in the ground state versus
$\omega_c$  and $g$ under the condition $\omega_c \omega_q=1$.
(a) The energy gap between the first excited states and the ground states.
(b) The occupation of the cavity field $\langle \hat{a}^{\dag}\hat{a}\rangle$.
(c) The quantum Fisher information of the cavity field defined in Eq.~(\ref{QFI}).
(d) The partial transposed criteria for entanglement.
(e) The mutual information.
(f) The minimum variance of the cavity field quadrature. }
\label{FigPhaseDiagram}
\end{figure}
\section{The Phase Diagram}\label{PhaseDiagram}
The critical coupling strength for the quantum phase transition of the Hamiltonia
(\ref{Eq:Hamiltonian}) is determined to be
\begin{equation}
g_{c} = \frac{\sqrt{\omega_c\omega_q}}{2}.
\label{Eq:CCS}
\end{equation}
To characterize the quantum phase transition of this model, we examine the behavior of
several quantities in the ground state. In addition to the excitations in the cavity
traditionally, we check the energy gap between the ground state and the first excited
state, the quantum Fisher information (QFI), the partial transposed criteria for
entanglement, the mutual information, and the fluctuation of the quadrature of the cavity
field. These diverse physical quantities serve as indicators for observing and analyzing
the phase transition within the scope of this work.

As $\omega_c\rightarrow0$ under the condition $\omega_c \omega_q=1$, the quantities in the
ground state tend to be non-analytical at the critical point as shown in Fig.\ref{FigPhaseDiagram},
which supports a second-order phase transition at zero temperature. As an example, the
excitation in the cavity field plays the role as the order parameter being zero in the
normal phase while acquiring positive values in the superradiance phase as shown in
Fig.\ref{FigPhaseDiagram} (b).

QFI is a fundamental concept in quantum metrology that stems from the classical Fisher
information~\cite{JSP1231}. It plays a crucial role in quantifying the sensitivity of a
quantum state used in parameter estimation, encompassing the critical factors of quantum
superposition and entanglement. To improve the precision of parameter estimation
in semi-classical Rabi model, there are works generalize the approximation expression
for maximal QFI in two-level system~\cite{EPJD72145}. The symmetric aspect of the the
rotating-wave and counterrotating-wave terms are discussed in the Rabi model~\cite{SCPMA66250313}.
QFI have been investigated for the atom in Jaynes-Cummings model coupling with the Ohmic
reservoir~\cite{IJTP593600}. These works manifest QFI has been investigated in various
aspects in order to benefit parameter estimation. Higher QFI indicates enhanced
sensitivity of the quantum system to fluctuations in the measured parameter. We focus on
evaluating the QFI pertaining to the state of the cavity field $\hat{\rho}_{c}$ with
respect to $\hat{\rho}_{c}(\theta)=e^{i\theta\hat{G}}\hat{\rho}_{c}e^{-i\theta\hat{G}}$,
where $\theta$ is the parameter need to be estimated as accurately as possible with
respect to the phase-shift generator $\hat{G}$, which depends on the target in certain investigation~\cite{PRL102100401,PRL106153603,PRA88043832}. The QFI reads
\begin{equation}
F=4\sum_{n}p_{n}(\Delta \hat{G})_{n}^{2}-\sum_{m\neq n}\frac{8p_{m}p_{n}}{
p_{m}+p_{n}}|\langle \psi _{m}|\hat{G}|\psi _{n}\rangle |^{2}, \label{QFI}
\end{equation}
where $\hat{\rho}_{c}|\psi_{n}\rangle$=$p_{n}|\psi_{n}\rangle$.
The first term of Eq.~(\ref{QFI}) is an expectation for each pure state
$|\psi_{n}\rangle$ with
$(\Delta \hat{G})_{n}^{2}\equiv\langle\psi_{n}|\hat{G}^{2}|\psi_{n}\rangle-|\langle\psi_{n}|\hat{G}|\psi_{n}\rangle|^{2}$.
The second term denotes the negative correction. Here, the QFI provides a quantitative
measure for the precision attainable in estimating the parameter $\theta$ in experiments
conducted on the quantum state. While states with larger QFI are indeed valuable for a
single-mode linear interferometer and can be treated as a kind of variance, we go beyond
their practical utility and view QFI as a witness for the phase transition in this work.
When $\frac{\omega_c}{\omega_q}$ is sufficiently small and $\omega_c \omega_q=1$, the
critical coupling strength equals 0.5. As shown in Fig.~\ref{FigPhaseDiagram}(c), the
behavior of QFI versus the coupling strength $g$ and $\omega_c$ coincides with the phase
diagram reflected by the energy gap between the ground state and the first excited state,
and that by $\langle \hat{a}^{\dag}\hat{a}\rangle$ as shown in Fig.\ref{FigPhaseDiagram}
(a) and (b), respectively. The eigen-energy degeneracy coincides with a spontaneous breaking
of the $Z_2$ parity symmetry which provides a fundamental view for this phase transition.

The partial transposed criterion provides a methodology for measuring entanglement in
bipartite quantum systems~\cite{PRL771413,PRA65032314}. It involves checking the eigenvalues
of the partially transposed density matrix of the hybrid system. The presence of negative
eigenvalues of the partially transposed density matrix means the presence of entanglement.
To capture this phenomenon and serve as a witness for the phase transition in this work,
we employ the absolute value of the summation of the negative eigenvalues $\lambda_i$ of
the partially transposed density matrix to witness the phase transition:
$|E^T|=\sum_{\lambda_i<0}|\lambda_{i}|$, in this work. As shown in Fig.~\ref{FigPhaseDiagram}(d),
the phase diagram indicated by $|E^T|$ coincides with those indexed by the energy gap between
the ground state and the first excited state, $\langle \hat{a}^{\dag}\hat{a}\rangle$, and the
QFI. However, it is crucial to acknowledge that the partial transpose criterion can only serve
as a sufficient condition for entanglement since there exist entangled states that maintain
a positive-definite nature after undergoing partial transposition~\cite{PRL771413,PRA65032314}.
This means the superradiance phase offers the source of entangled states which is useful in
quantum information process.

Mutual information (von Neumann mutual information) is another quantity demonstrating correlation
between the subsystems. It characterizes the information of sub-system `$A$' by exploring its
counterpart `$B$'~\cite{RMP80517,PRA91012301,PRL90050401}.
It is defined as $I^M=S_A+S_B-S_{AB}$, where $S_{A(B)}$ is the entropy for $A(B)$ system
and $S_{AB}$ is that for the hybrid bipartite system~\cite{Shannon}.  The entropy can be calculated
as $S(t)=-Tr[\rho(t)log(\rho(t))]$  where $Tr[\bullet]$ denotes the trace of $\bullet$. Although
its definition is distinct from the entanglement witness $|E^T|$ mentioned above, their behavior
is quite similar in indicating the phase transition in this work as shown in
Fig.~\ref{FigPhaseDiagram}(d) and (e). Entanglement and mutual information provide
quantum resource in quantum information process. This inspires us seeking quantum source in
other quantum phase transitions.

In general, it is believed that there is a relationship between phase transitions and fluctuations
in physical systems~\cite{Sachdev,RPP662069}. To accurately capture the phase transitions in the
strong and ultra-strong coupling regime by using quadrature measurements, it is crucial to define
positive and negative frequency cavity-photon operators as
$\hat X^{+}=\sum_{j,k>j}X_{jk}|j\rangle \langle k|$ and $\hat X^{-}=(\hat X^{+})^{\dagger}$,
with $X_{jk}\equiv \langle j|\hat a^{\dagger}+\hat a|k\rangle$, in the dressed
eigen-basis $|j\rangle$, $|k\rangle$ of the Hamiltonian (\ref{Eq:Hamiltonian}) with
eigen-values $\omega_j$ and $\omega_k$, respectively~\cite{PRA88063829,PRA92063830}. This
step is essential for excluding unphysical streams of output photons in experiments. In
the limit of weak coupling, these operators coincide with $\hat{a}$ and $\hat{a}^\dag$,
respectively. And the similar operators can be defined for $\hat \sigma_-$ and
$\hat \sigma_+$~\cite{PRA88063829,PRA92063830}. Using the defined operators above, we
can assess the minimum variance of the quadrature of the cavity field defined as
$\hat{X}(\theta) = \frac{\hat{X}^{-} e^{i\theta} + \hat{X}^{+} e^{-i\theta}}{\sqrt{2}}$,
with $\theta \in [0,2\pi)$. The minimum variance reads
$V_{m}^{a}=\langle\hat{X}^2(\theta_m)\rangle-\langle \hat{X}(\theta_m)\rangle ^{2}$ versus
$\theta_m$. The landscape of $V_{m}^{a}$ versus $g$ and $\omega_c$ coincides with those of
the quantities used above by comparing Fig.~\ref{FigPhaseDiagram}(f) to the other panels
Fig.~\ref{FigPhaseDiagram}. The behaviors of all these quantities approaching the critical
point indicate that the smaller $\frac{\omega_c}{\omega_q}$, the more obvious of the phase
transition. Exploring these quantities not only provides additional avenues for
studying phase transitions in experiments, but also allows us to understand the essence of
phase transitions from multiple perspectives.
\begin{figure}[htbp]
\includegraphics[width=\linewidth]{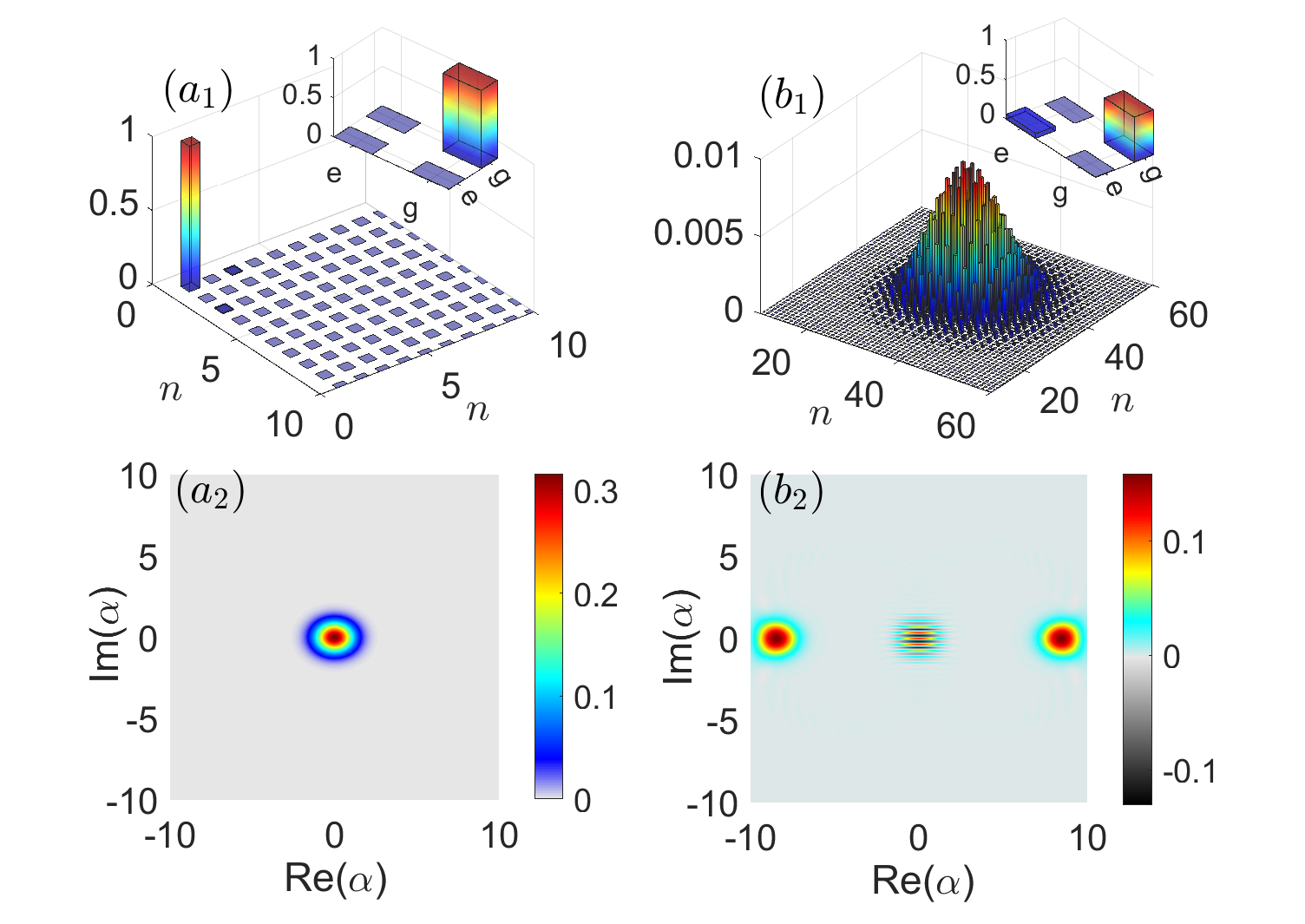}%
\caption{$(a_1)$ and $(a_2)$ show the ground state and Wigner function of cavity mode when
coupling $g=0.2$ and $(b_1)$. $(b_2)$ are those when coupling $g=0.7$ when $\omega_c=0.1$.
The insets in $(a_1)$ and $(a_2)$ show the corresponding ground state of the qubit.}
~\label{Figas3phaseW}
\end{figure}
To gain insight into the phase transition, we examine the density matrices of the atom,
cavity field, and the Wigner function of the cavity mode in the ground states in
different phase regions. It can be seen in Fig.~\ref{Figas3phaseW}, in the normal phase,
there is no excitation in both the cavity and the atom (qubit). The Wigner function is
that of the vacuum state with Gaussian distribution as shown in Fig.~\ref{Figas3phaseW}
$(a_2)$. However, upon entering the strong coupling regime, a distinctive feature emerges
in the Wigner function. Negative texture between the two peaks indicate a Schr$\ddot{o}$dinger
cat state, a nonclassical resource which can be used in quantum information process~\cite{PRA96013824}.
For the condition $\omega_c/\omega_q=0.1$, one excitation of qubit is equivalent to ten
excitation of the cavity field. This leads to the low level of the excitation of the qubit
in the superradiance phase as shown in the inset of Fig.~\ref{Figas3phaseW} $(b_1)$.
\section{Dynamics Reflect Phase Transition}
\label{Dynamics}
In this work, we put forward a conjecture that the dynamics of the quantities (include the
above quantities) can reveal the quantum phase transition. To verify this idea, we check
the dynamics of the quantities in terms of whether the coupling strength quenches across
the critical point or not.

Before we check the dynamics in quench, we know that complete isolation of a quantum system
from its environment remains a formidable challenge. For the open quantum system, we resort
to numerical calculations by solving the master equation under the inherent damping effects
of both the cavity field and qubit. However, we implement the finite size of a 50-photon
Hilbert space which imposes certain limits on the accuracy of our numerical calculations.

In the strong and ultrastrong coupling regimes, to exclude unphysical counting of photons,
an effective approach for describing the system involves solving the master equations in the
eigen-basis of the Hamiltonian (\ref{Eq:Hamiltonian}). In this case, the master equation reads
\begin{equation}\label{eq:master-eq}
\dot\rho(t) = i [\rho(t), \hat{H}] + \mathcal{L}_{\hat{a}}\rho(t) + \mathcal{L}_{\hat{\sigma}^{-}}\rho(t),
\end{equation}
where $\mathcal{L}_{\hat{a}}$ and $\mathcal{L}_{\hat{\sigma}^{-}}$ are Liouvillian superoperators
describing the decoherence of the cavity field and qubit~\cite{PRA84043832}. They read
$ \mathcal{L}_{\hat{x}}\rho(t) = \sum_{j,k>j}\Gamma^{j k}_{\hat{x}}\bar{n}(\Delta_{k j},T)\mathcal{D}[|k \rangle \langle j|]\rho(t) + \sum_{j,k>j}\Gamma^{j k}_{\hat{x}}(1 + \bar{n}(\Delta_{k j},T))\mathcal{D}[|j \rangle \langle k|]\rho(t)$ for $\hat{x} = \hat{a}, \hat{\sigma}^{-}$ with $\mathcal{D}[\mathcal{O}]\rho(t)$ =
$\frac{1}{2} (2 \mathcal{O}\rho(t)\mathcal{O}^{\dagger}-\rho(t) \mathcal{O}^{\dagger} \mathcal{O} - \mathcal{O}^{\dagger} \mathcal{O}\rho(t))$.
The relaxation coefficients $\Gamma^{j k}_{\hat{x}} = 2\pi d(\Delta_{k j}) \alpha^{2}_{\hat{x}}(\Delta_{k j})| C^{\hat{x}}_{j k}|^2$ with $d(\Delta_{k j})$ being the spectral density of the baths, $\alpha_{\hat{x}}(\Delta_{k j})$ denoting
the system-bath coupling strength, $\Delta_{k j} = \omega_{k} - \omega_{j}$, and
$C_{j k}^{\hat{x}} = -i \langle j |(\hat{x} - \hat{x}^{\dagger})| k \rangle$ .
$\bar{n}(\Delta_{k j},T)=[\exp(\Delta_{k j}/T)-1]^{-1}$ is the mean number of quanta in a mode with
frequency $\Delta_{k j}$ and temperature $T$ (Boltzmann constant $k_B$=1). When considering a cavity
coupling to the momentum quadrature of a field in one-dimension waveguides, the spectral density
$d(\Delta_{k j})$ is constant and $\alpha_{\hat{x}}^{2}(\Delta_{k j}) \propto \Delta_{k j}$. Then
the relaxation coefficients reduce to
$\Gamma^{j k}_{\hat{x}} = \gamma_{c} \,(\Delta_{k j}/\omega_{0}) \, |C^{\hat{x}}_{j k}|^2$
where $\gamma_{c}$ is the standard damping rate. These assumptions can be realized in circuit-QED~\cite{PRX2021007}
or trapped ion system~\cite{PRL118073001,PRX8021027,NC121126}.
The influence of dephasing and Lamb shifts in current experiments can be deemed negligible as they
do not exert significant influence~\cite{PRL118073001,PRX8021027,NC121126}.

\subsection{Dynamics in Quench}
\label{DPT}
\begin{figure}[htbp]
\includegraphics[width=\linewidth]{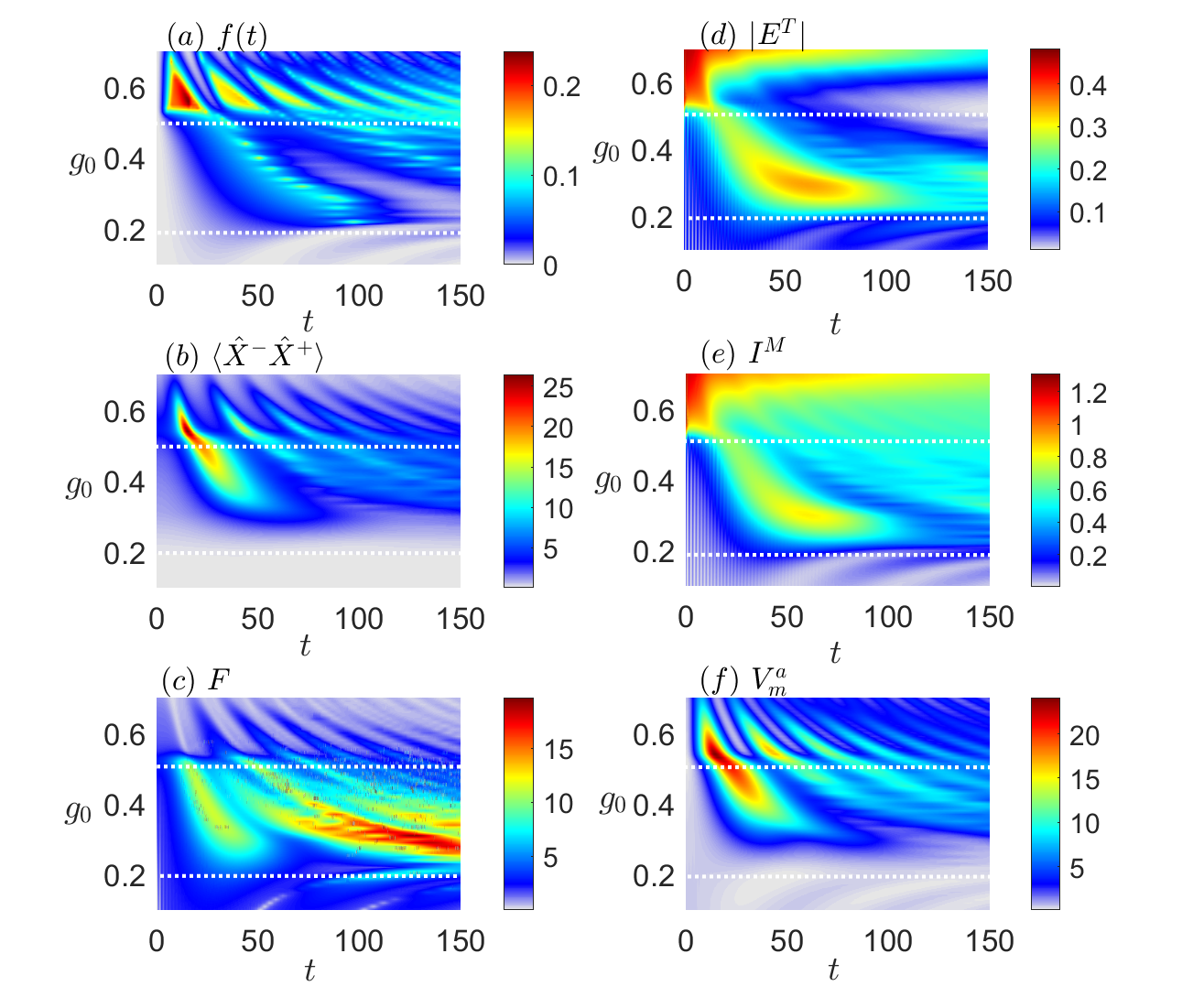}
\caption{The dynamics of the quantities when the coupling strength $g$ changes
suddenly from $g_0$ to $g^{\prime}=g_0+0.3$. $(a)$ The return rate $f(t)$.
$(b)$ $\langle \hat{X}^{-}\hat{X}^{+}\rangle$. $(c)$ The quantum Fisher information.
$(d)$ The absolute value of the of the summation of the negative eigenvalues of
the partial transposed density matrix. $(e)$ The mutual information.
$(f)$ The minimum variance of the cavity field quadrature.
$\gamma_{c}$=$\gamma_{q}$=0.01$\omega_{c}$.}
\label{fig:L_DQPT}
\end{figure}
The concept of dynamical quantum phase transition has emerged from the analogy
between an equilibrium partition function and the return probability in many-body
unitary dynamics~\cite{PRL110135704}. This expansion of criticality to non-stationary
scenarios involves sudden changes in the macroscopic properties of quantum systems 
over time. Dynamical phase transition in quantum systems is usually investiated by 
quench ~\cite{PRL110135704,PRB102060409,PRB91155127}.
However, quench is not limited to investigate dynamical quantum phase transition. 
The behavior of physical quantities in quench can be used to reveal other intriguing 
physics. In this work, we would check the behaviors of the physical quantities 
mentioned above in quench to examine whether they can be used to reveal the quantum 
phase transition. The return rate is usually employed in exploring dynamical phase 
transition $f(t)=-\frac{1}{N} \ln G(t)$~\cite{PRL110135704,PRB102060409,PRB91155127}.
This quantity behaves non-analytically versus time when the Loschmidt overlap
$G(t)$ vanishes. Here the Loschmidt overlap $G(t)$ is defined as
$G(t)=|\langle \psi_{g_0;0} | e^{-i H^{\prime} t} |\psi_{g_0;0} \rangle|$.
$|\psi_{g_0;0}\rangle$ denotes the initial state with coupling strength $g_0$ in
the Hamiltonian. This quantity measures the overlap between the time evolved state
$e^{-i H^{\prime} t} |\psi_{g_0;0} \rangle$ and the initial state $|\psi_{g_0;0}\rangle$
following a sudden change of the parameter $g$ in the post-quench Hamiltonian, namely,
$H(g_0)\underrightarrow{~quench~}H(g^{\prime})$ in this work. To employ this quantity
reflecting the phase transition, we check the behavior of $f(t)$ as shown
in Fig.~\ref{fig:L_DQPT}(a) when the coupling strength changes suddenly from $g_0$ to
$g^{\prime}=g_0+0.3$. It can be seen that the cusps appear when the coupling strength
quenches across the critical point, namely $g_0\in[0.2,0.5]$.

Fig.~\ref{fig:L_DQPT}(b)-(f) show the dynamics of $\langle \hat{X}^{-}\hat{X}^{+}\rangle$,
QFI, entanglement witness $|E^{T}|$, mutual information $I^{M}$
and the minimum variance $V^{a}_m$ versus the initial coupling strength $g_0$ in the
quench $H(g_0)\underrightarrow{~quench~}H(g_0+0.3)$. In this manner of quench, there
are three parameter regions, namely $g_0<0.2$; $g_0\in[0.2,0.5]$; $g_0>0.5$, with
distinct dynamical characters. It depends on whether the coupling strength $g$ quench
across the critical point, namely $g_c=0.5$, or not for all these quantities. It is
interesting to notice that the behavior of $\langle \hat{X}^{-}\hat{X}^{+}\rangle$ and
$V^{a}_m$ are similar when $g_0<g_c$. $|E^{T}|$ and $I^{M}$ behave similarly versus $g_0$
although their formalism are different obviously. This hints resemblance between these
quantities. These characters of dynamics provide avenues to check such a phase transition
in experiments by observing the dynamics of physical quantities. And it suggests a
dynamical manner to obtain quantum resources.
\begin{figure}[htbp]
\includegraphics[width=\linewidth]{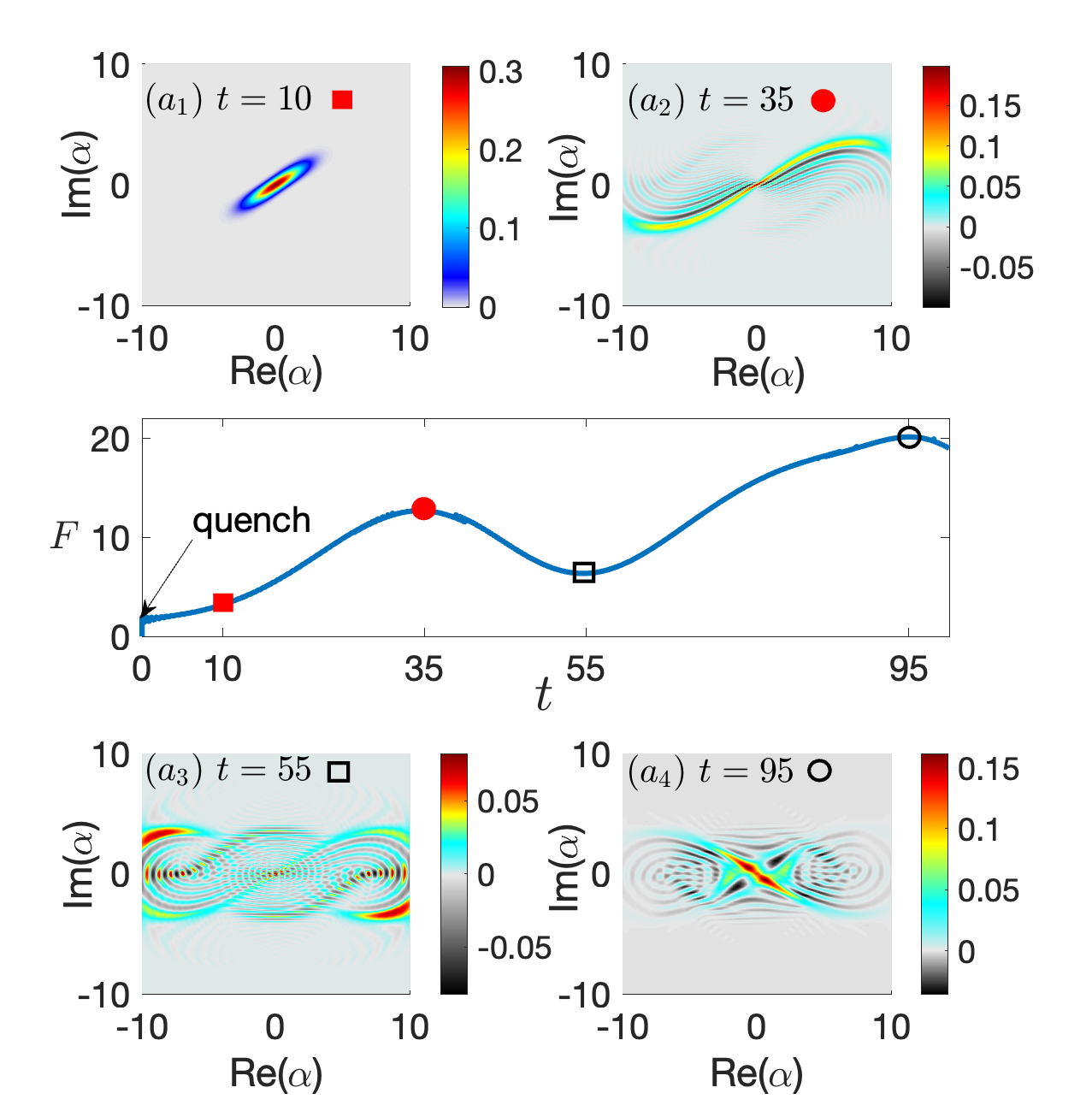}
\caption{The dynamics of the QFI when the coupling strength $g$ changes suddenly from
$g_0=0.35$ to $g^{\prime}=0.65$ at temperature $T=0$. $(a_1)$-$(a_4)$ are the Wigner
functions of the cavity mode at $t$=10, 35, 55, and 95, respectively.
$\gamma_{c}$=$\gamma_{q}$=0.01$\omega_{c}$. The other parameters are same to those in
Fig. \ref{fig:L_DQPT}.}
\label{fig:squeezM}
\end{figure}

To gain insight into the dynamics of the states during the quench, we check examples
of the Wigner function for states when $g_0=0.35$ changes to $g^{\prime}=0.65$ abruptly
in Fig.~\ref{fig:squeezM} at zero temperature. Negative scars in the Wigner function
indicate the states being non-classical starting from the vacuum. Such nonclassical
states prefer redundancy encoding in quantum information~\cite{PRA96013824}.
\begin{figure}[htbp]
	\includegraphics[width=\linewidth]{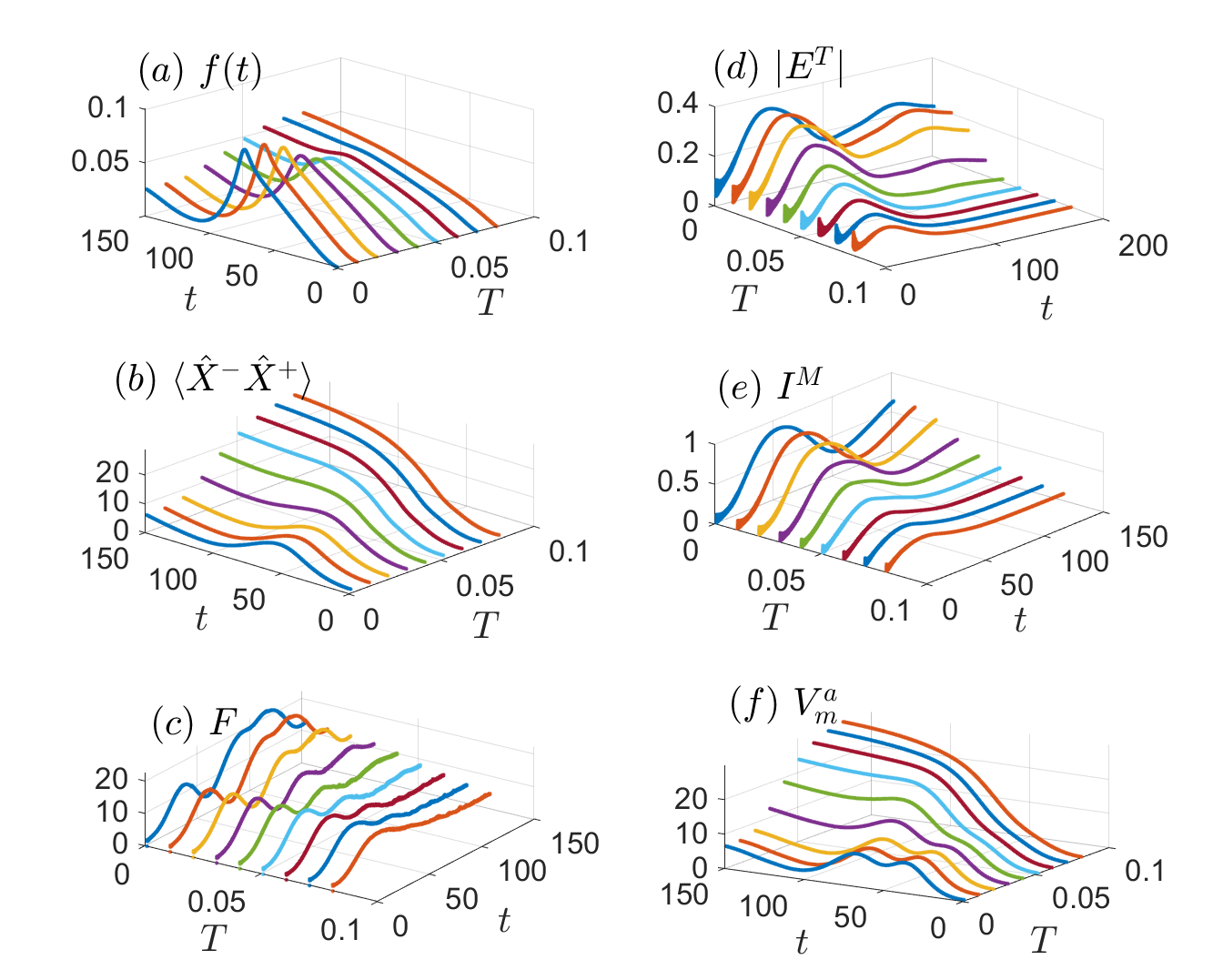}
	\caption{The dynamics of the quantities when the coupling strength $g$ changes
suddenly from 0.35 to 0.65 versus temperature $T$. The other parameters
are same to those in Fig.~\ref{fig:L_DQPT}.}\label{fig:L_DQPT_Quench}
\end{figure}
\section{Influence of Temperature of The Environment}
\label{InfluenceT}
In the results above, zero-temperature environment is assumed. However the
temperature is indeed a factor which should be considered when cavity interacting
with atoms~\cite{JPC1234035}. It is necessary to examine the influence of temperature
on the dynamics. In Fig.~\ref{fig:L_DQPT_Quench}, we plot the dynamics of the quantities
mentioned in Sec.~\ref{Dynamics} above versus the temperature. It is clear to see that
with increasing of the temperature $T$, the characteristic bends in $f(t)$,
$\langle \hat{X}^{-}\hat{X}^{+}\rangle$, QFI, and variance $V^{a}_m$ and correlations
reflected by $|E^T|$ and $I^M$ during evolution tend to disappear and becomes gentle
for these quantities. According to the master Eq.~(\ref{eq:master-eq}), higher temperature
leads to more intense incoherent dissipation and driven. These are the negative factors
for quantum resource and the application of these quantities as the indicator for the
phase transition. This means low temperature benefits revealing the phase transition by
the dynamics.
\section{Potential Experimental Systems}
\label{Experiments}
In this work, we consider the phase transition of the Rabi model with the coupling
strength vary across weak and strong regime and reveal the phase transition by the dynamics
of several quantities. There are different quantum platforms to check these results. For
instance, within a single trapped ion system, it is feasible to adjust the coupling strength
between the atom and the Boson mode to achieve the Rabi model. This provides an opportunity
to investigate the phase transitions and the dynamics in the ultrastrong and deep strong-coupling
regimes, overcoming the limitations of the rotating-wave approximation~\cite{PRX8021027,PRL118073001}.
The entanglement and correlation between the Boson mode and the two-level system can be detected
in such models. A similar phase transition can be illustrated by employing a $^{171}$Yb+ion
within a Paul trap through the adiabatic adjustment of the coupling strength between the ion
and its spatial mode, without any thermodynamic constraints~\cite{NC121126}. In addition to
trapped ion systems, the strong-coupling regime can also be achieved in circuit quantum
electrodynamics setups, where superconducting artificial atoms are coupled to on-chip
cavities~\cite{NP6772} or coupled to the electromagnetic continuum of a one-dimensional
waveguide~\cite{NP1339}. These systems offer the tunability that expands the capabilities
of quantum optics, enabling exciting investigations into ultrastrong interactions between
light and matter. Yet, the quantum Rabi model in the ultra-strong coupling regime can
be realized by a superconducting circuit embedded in a cavity QED setup~\cite{NC8779}. Through
the coupling of a flux qubit and an LC oscillator using Josephson junctions, it is possible
to realize the circuits Rabi model with a wide range of coupling strengths~\cite{NP1344}. This
enables the exploration of the ground-state phase transition and entanglement mentioned in our
work. Quantum simulation offers various avenues to realize the Rabi model in diverse systems.
One proposal is using a circuit quantum electrodynamics chip with moderate coupling between a
resonator and transmon qubit to achieve precise digital quantum simulation of deep-strong coupling
dynamics~\cite{NC81715}. This proposal will enable exploration of extreme coupling regimes and
quantum phase transitions as mentioned in our work. A practical implementation of a photonic
analog simulator for the quantum Rabi model has been achieved using femtosecond laser-written
waveguide superlattices. This advancement offers a tangible experimental platform for investigating
the intricate physics of light-matter interaction in the deep strong coupling regime~\cite{PRL108163601}.
In these potential platforms, tuning the coupling quickly is necessary to realize quench.

\section{Conclusion}\label{Conclusion}
In addition to traditional indicators such as oscillator occupation and qubit excitation,
we first show that the quantum Fisher information, qubit-oscillator entanglement, mutual
information, and the variance of the cavity field quadrature display minimal values below
the transition point. However, when the coupling strength is tuned across the quantum
critical point, these quantities undergo swift and substantial growth. This transition
is attributed to the degeneracy between the ground state and the first excited state.
The physical quantities display singular behavior when the ratio of the frequencies of
oscillator and the qubit approaches zero. This behavior is analogous to approaching closer
to a thermodynamic limit in superradiant phase transition. Nonclassical Schr\"{o}dinger
cat state is revealed by Wigner function in the superradiant phase. Then we examined the
dynamics of the Rabi model to investigate the probability of indicating the quantum phase
transition by dynamics. The quantities used to witness the phase transition in the ground
state all behave differently depend on whether it quenches across the critical point. It
offers avenues to reveal the phase transition by quantities in non-equilibrium process.
And it is shown that temperature is poisonous to applying this quench method result from
the incoherent process. There are platforms can be considered to realize our work, for
example, the trapped ion system, circuit quantum electrodynamics setups like superconducting
artificial atoms coupled Boson modes, quantum simulation using circuit quantum electrodynamics
chip or femtosecond laser-written waveguide superlattices. These systems allow the researchers
to vary the experimental parameters and study their influence on the phase transition.

\begin{acknowledgments}
X. L. Zhao thanks H. J. Xing for helpful discussions and National Natural
Science Foundation of China, No.12005110 and Natural Science Foundation
of Shandong Province, China, No.ZR2020QA078.
\end{acknowledgments}

\end{document}